\documentclass[prl,final,twocolumn,showpacs,showkeys,superscriptaddress]{revtex4}
\usepackage{amsmath}
\usepackage{graphicx}
\usepackage{amsfonts}
\usepackage{amssymb}
\begin{document}

\title{Spectroscopy of Three-Particle Entanglement in a Macroscopic Superconducting Circuit}

\author{Huizhong Xu}
\author{Frederick W. Strauch}
\author{S. K. Dutta}
\author{Philip R. Johnson}
\author{R. C. Ramos}
\affiliation{Department of Physics, University of Maryland, College Park, Maryland 20742-4111}
\author{A. J. Berkley}
\affiliation{Department of Physics, University of Maryland, College Park, Maryland 20742-4111}
\affiliation{D-Wave Systems Inc., Vancouver, BC, Canada V6J4Y3}
\author{H. Paik}
\author{J. R. Anderson}
\author{A. J. Dragt}
\author{C. J. Lobb}
\author{F. C. Wellstood}
\email[Electronic address: ]{well@squid.umd.edu}
\affiliation{Department of Physics, University of Maryland, College Park, Maryland 20742-4111}

\date{\today}

\begin{abstract}
We study the quantum mechanical behavior of a macroscopic, three-body, superconducting circuit. Microwave spectroscopy on our system, a resonator coupling two large Josephson junctions, produced complex energy spectra well explained by quantum theory over a large frequency range. By tuning each junction separately into resonance with the resonator, we first observe strong coupling between each junction and the resonator. Bringing both junctions together into resonance with the resonator, we find spectroscopic evidence for entanglement between all three degrees of freedom, and suggest a new method for controllable coupling of distant qubits, a key step toward quantum computation.
\end{abstract} 
\maketitle
The main promise of using solid-state devices for quantum computation \cite{Nielsen2000} is that it will be relatively easy to scale such a technology from an individual qubit to the large number of qubits ultimately required for key applications \cite{Shor94}. A variety of individual qubits based on superconducting devices \cite{Makhlin2001} have been implemented \cite{Nakamura99,Martinis2002}. Work has also been reported on entangled states in two coupled charge qubits \cite{Pashkin2003}, Josephson-junction phase qubits \cite{Berkley2003s}, flux qubits \cite{Izmalkov2004}, and most recently the coherent dynamics of a flux qubit coupled to its SQUID detector \cite{Chiorescu2004}. The next challenge for scaling is to produce the multiparticle entangled states needed for error correction \cite{Shor95} and teleportation \cite{Bennett93}, preferably in a device that controllably couples distant qubits.

\begin{figure}[b]
\begin{center}
\includegraphics{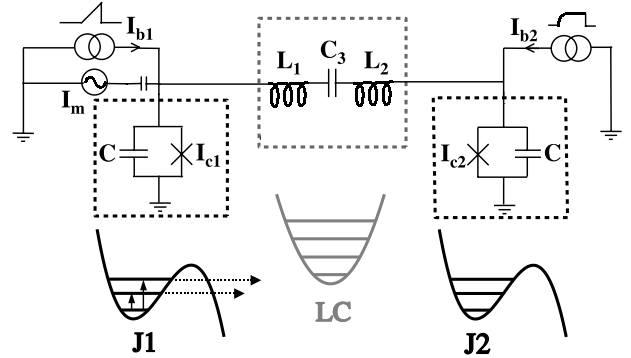}
\caption{Schematic of a macroscopic superconducting three-body system which consists of two Josephson junctions and an LC resonator. Parameters obtained from fitting the spectra in Fig. 2 are: $I_{c1} = 21.388 \pm 0.003\mu$A, $I_{c2} = 22.536 \pm 0.003\mu$A, $C = 4.85 \pm 0.05$ pF, $L \equiv L_1 + L_2 = 1.7 \pm 0.05$ nH and $C_3 = 0.33 \pm 0.01$ pF. The two junctions J1 and J2 (in the left and right dashed boxes) have anharmonic potentials with unequal energy level spacings (shown below). The horizontal arrows represent quantum tunneling and the vertical arrows denote microwave-induced transitions. The LC oscillator (in the center dashed box) has a harmonic well with equal energy level spacings.}
\label{circuit}
\end{center}
\end{figure}

A new approach to the scaling of superconducting qubits \cite{Blais2004} utilizes an analogy to the strong-coupling regime of atomic cavity-QED experiments \cite{Raimond2001}.  This analogy was recently realized in an elegant experiment \cite{Wallraff2004}, in which a single Cooper-pair box qubit (the atom) was capacitively coupled to a superconducting transmission line (the cavity).  The sub-$\mu$m sized charge qubit was first characterized by measurements of the resonator in the dispersive regime.  This was followed by the observation of the resonant vacuum Rabi splitting, a spectroscopic indication of entanglement between the charge qubit and a single photon in the resonator.   

We present experimental results that extend this new field of superconducting cavity-QED to three macroscopic qubits---two Josephson junctions and a resonator, the analog of two atoms and a cavity.   Figure 1 shows a circuit schematic of our system, which consists of two large ($10 \mu\text{m} \times 10 \mu\text{m}$) Josephson-junction phase qubits connected together by a series inductor-capacitor (LC) resonator.  This system is distinct from atomic cavity-QED systems, in that our ``atoms'' are distinguishable and independently tunable. We first use spectroscopic measurements to study the coupling of each junction to the LC oscillator. We then couple \emph{all three} degrees of freedom together, and observe spectroscopic evidence in clear agreement with quantum mechanics.

The three degrees of freedom of this system are the macroscopic quantum variables  $\gamma_1$ and $\gamma_2$ (the gauge-invariant phase differences across junctions J1 and J2, respectively), and $\gamma_3 = 2 \pi L I/\Phi_0$ corresponding to the current $I$ flowing through the total inductance $L=L_1+L_2$.  Each degree of freedom corresponds to distinct coherent motions of billions of electron pairs, and is therefore macroscopic both in size and number \cite{Leggett80}.  Using standard circuit analysis, the Hamiltonian for the system is

\begin{widetext}
\begin{equation}
H =  \overset{H_{\text{J1}}}{\overbrace{\frac{p_1^2}{2m} - \frac{\Phi_0}{2\pi}(I_{c1} \cos \gamma_1 + I_{b1} \gamma_1)}} + \overset{H_{\text{J2}}}{\overbrace{\frac{p_2^2}{2m} - \frac{\Phi_0}{2\pi}(I_{c2} \cos \gamma_2 + I_{b2} \gamma_2 )}} + \overset{H_{\text{LC}}}{\overbrace{\frac{p_3^2}{2m_3} + \frac{1}{2}m_3 \omega_3^2 \gamma_3^2}} + \overset{H_{coupling}}{\overbrace{\xi \frac{p_1^{\,} p_3}{\sqrt{m m_3}} - \xi \frac{p_2^{\,} p_3}{\sqrt{m m_3}}}}
\end{equation}
\end{widetext}
where $p_i$ ($i=1,2,3$) are the canonical momenta of the three degrees of freedom with corresponding effective masses of $m_1=m_2=m=C (\Phi_0/2\pi)^2$ and $m_3 = (\Phi_0/2\pi)^2 C_3 C/(C+2C_3)$. The quantity $\Phi_0 = h/2e$ is the flux quantum, $C$ is the junction capacitance for J1 and J2, $C_3$ is the capacitance of the LC resonator, $I_{c1}$ and $I_{c2}$ are the junctions' critical currents, $I_{b1}$ and $I_{b2}$ are two steady bias currents,  $\omega_3 = 1/\sqrt{L C_3 C/(C+2 C_3)}$ is the angular frequency of the LC resonator, and $\xi = \sqrt{C_3/(C+2 C_3)}$ is a dimensionless coupling coefficient.  

The first term in $H$, $H_{\text{J1}}$, is the Hamiltonian for J1 alone.  It has dynamics analogous to that of a particle moving in a tilted washboard potential (see Fig.~1). Metastable energy states \cite{Martinis85,Berkley2003B} are present in the well and can be observed if the qubit is well isolated. The potential and the level spacings can be controlled by the bias current $I_{b1}$. The metastable states have zero dc voltage, but can tunnel \cite{Voss81} to continuum states that exhibit a finite dc voltage across the junction. We can probe the states by applying microwave current $I_m$ that can drive transitions from the ground state to the excited states.  These excited states have much higher tunneling rates and thus are easily detected.

The second term, $H_{\text{J2}}$, describes J2, which has dynamics similar to J1 but is independently controlled by its bias current $I_{b2}$.  $H_{\text{LC}}$ describes the harmonic oscillator dynamics of the LC resonator (see Fig.~1).  Finally, $H_{coupling}$ represents the capacitive coupling of each junction to the resonator.  Note that the momenta $p_i$ are proportional to the charges stored on each capacitor in the circuit \cite{Johnson2003}, and thus the coupling is electrostatic. 

Our Josephson junctions are thin-film $10 \mu\text{m} \times 10 \mu\text{m}$ Nb/AlOx/Nb junctions made by Hypres, Inc. on a $5 \text{mm} \times 5 \text{mm}$ silicon chip. The critical currents of the junctions are $\sim120 \mu$A in zero magnetic field, but can be adjusted by applying an external magnetic field. The coupling inductor is a $780 \mu\text{m} \times 90 \mu\text{m}$ thin-film niobium loop connecting the two junctions, and the coupling capacitance physically consists of two capacitors in series, each formed by $60 \mu\text{m} \times 60 \mu\text{m}$ parallel niobium plates separated by a 200 nm layer of $\text{SiO}_2$. With this geometry we estimate the inductance $L \approx 1.5$ nH  and the capacitance $C_3 \approx 0.37$ pF. The chip is mounted inside a Cu box which is attached to the mixing chamber of a dilution refrigerator with a base temperature of 25 mK. We note that this same chip was previously examined \cite{Berkley2003s} at junction frequencies less than $\omega_3$ (with $I_{c1} \sim I_{c2} \sim 15 \mu$A), where the effect of the LC resonator reduced to purely capacitive coupling.  For the higher frequencies considered here ($I_{c1} \sim I_{c2} \sim 22 \mu$A), however, its effect should be described by the Hamiltonian in Eq. (1).

\begin{figure*}[t]
\begin{center}
\includegraphics{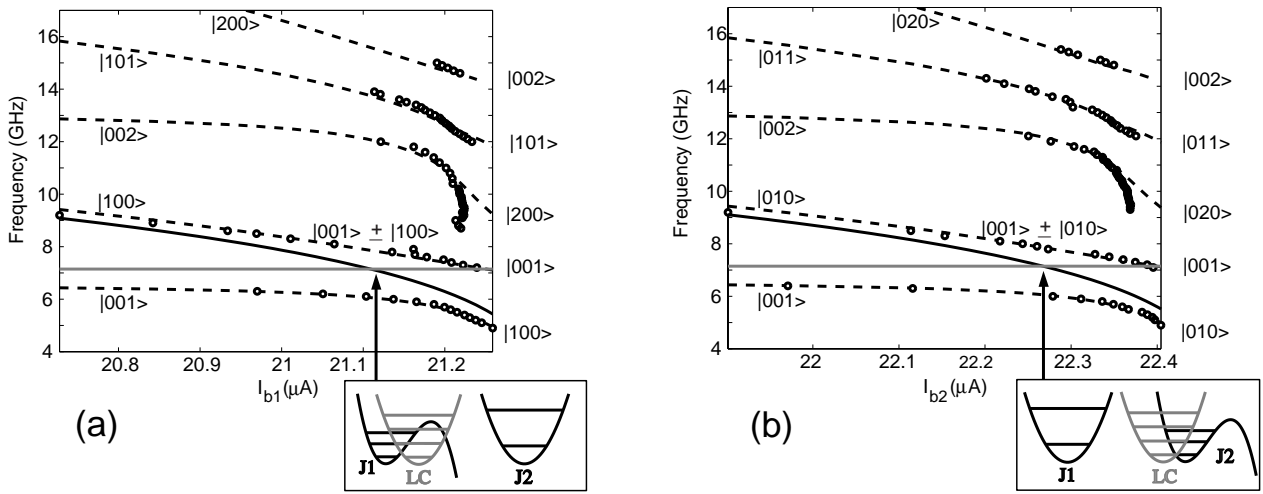}
\caption{(a) Spectrum of the system when the bias current $I_{b1}$ for junction J1 is ramped and $I_{b2}=0$ for junction J2. Circles are measured microwave resonance positions, dashed black lines are from quantum mechanical calculations using Eq.~(1) with parameters given in Fig. 1 and $I_{b2}=0$. The solid curves denote uncoupled $|0\rangle$ to $|1\rangle$ level spacings for J1 (black) and the resonator (gray), while that for J2 ($\sim 19$ GHz) is not shown in the plot. For low and high bias, the energy level transitions are from the ground state $|000\rangle$ to the indicated higher excited states, where the state notation is $|\text{J1, J2, LC} \rangle$. At the degeneracy point $I_{b1} \approx 21.12 \mu$A (shown in the lower box), the first two excited states are $(|001\rangle \pm |100\rangle)/\sqrt{2}$. The deviation of the fit for the third excited state is probably due to its large tunneling rates at high bias currents. (b) Spectrum of the system when J2 is ramped and J1 is zero-biased ($I_{b1}=0$). Similar to (a), there is a degeneracy point at $I_{b2} = 22.27 \mu$A (shown in the lower box) where the two states are $(|001\rangle \pm |010\rangle)/\sqrt{2}$.}
\label{jjlc}
\end{center}
\end{figure*}

Figure 2(a) shows the spectrum of the system when the bias current $I_{b1}$ for junction J1 is ramped and that for junction J2 is held at $I_{b2}=0$. The circles denote measured resonance peak positions when microwaves are continuously applied to induce transitions from the ground state to excited states. The dashed lines are from quantum mechanical calculations using the Hamiltonian in Eq. (1) with the parameters in Fig. 1 (obtained by fitting) and $I_{b2}=0$.  The zero-biased junction J2 is effectively decoupled from the rest of the system since it has a much larger energy scale ($\sim19$ GHz) than both junction J1 and the LC resonator ($\sim7$ GHz). Therefore, we observe a spectrum essentially due to junction J1 and the LC resonator only. The avoided crossing between the first and second excited states occurs at $I_{b1} \approx 21.12 \mu$A. States of the form $(|001\rangle \pm |100\rangle)/\sqrt{2}$ are predicted here, where the first, second, and third positions in the ket denote the states of J1, J2, and the LC oscillator, respectively.  The next three excited states at the degeneracy point are $(|002\rangle + |200\rangle - \sqrt{2}|101\rangle)/2$ , $(|002\rangle - |200\rangle)/\sqrt{2}$, and $(|002\rangle + |200\rangle + \sqrt{2}|101\rangle)/2$. We note these states are entangled only between junction J1 and the LC resonator, since junction J2 is frozen in its ground state. 

Figure 2(b) shows the measured spectrum for the case of $I_{b1}=0$ while ramping the bias current for J2. Similarly, comparison with theory reveals the states here describe entanglement between junction J2 and the LC resonator. We find good agreement between data and theoretical calculations using the same parameters as listed in Fig.~1 and $I_{b1}=0$.  

The observation of higher order transitions in Fig.~2, such as transitions from $|000\rangle$ to states involving $|002\rangle$ (analogous to a two-photon state in cavity-QED) provides strong evidence for the quantum nature of the system. Coupling the LC oscillator to the anharmonic junctions has introduced nonlinearity that allows us to distinguish these quantum transitions from the resonances of a classical harmonic oscillator. We also note that a single set of five parameters has been used to fit the ten curves in Fig.~2.  The good agreement between data and theory obtained here cannot be achieved by any classical model that includes only three degrees of freedom. Thus by tuning one junction into resonance with the LC resonator, we have observed spectroscopic evidence for entanglement analogous to the recent coherent coupling of a single Cooper-pair box to a superconducting transmission line.

We next show spectroscopic evidence for entangled states between \emph{two} junction qubits and an LC resonator. Figure 3 shows the measured spectrum when J2 is biased at a constant current and the bias current for J1 is ramped. Using the previously determined parameters, we compute the energy levels by adjusting the only remaining parameter $I_{b2} = 22.330 \mu$A.  That is, all six curves shown in Fig. 3 have been fit using just one parameter 
\footnote{{The measured $I_{b2}$ is $22.110 \mu$A; the discrepancy between this and the fitted value appears to come from the calibration of the current ramp which has overestimated the critical current of junction J2 by 1\%.}}. This clearly demonstrates that the multi-level spectroscopic measurements are well explained by the quantum mechanics of the Hamiltonian given in Eq. (1).  

\begin{figure}[t]
\begin{center}
\includegraphics{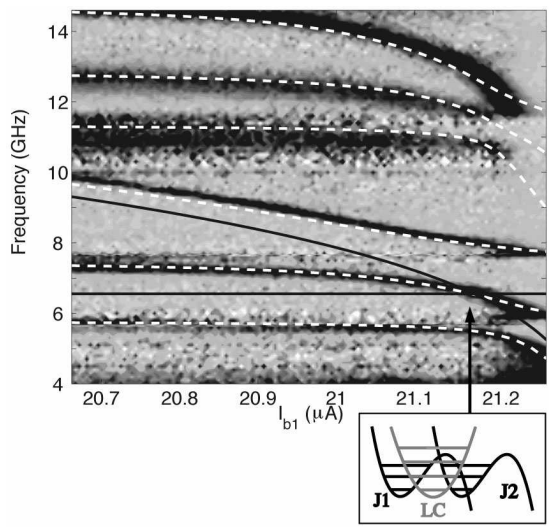}
\caption{Enhancement in escape rate when the bias current $I_{b1}$ for junction J1 is ramped and junction J2 is biased at a constant current of $I_{b2} = 22.330 \mu$A. Black corresponds to highest enhancement and light gray to zero enhancement. The white dashed lines are from quantum mechanical calculations using Eq.~(1) with parameters given in Fig.~1 and $I_{b2} = 22.330 \mu$A.  The solid black lines indicate the uncoupled $|0\rangle$ to $|1\rangle$ level spacings for J1 (curved) and J2 (horizontal), while that for the LC resonator ($\approx 7.1$ GHz) is not shown in the plot. At the triple degeneracy point $I_{b1} \approx 21.15 \mu$A (shown in the lower box), the three lowest excited states are $(|100\rangle - |010\rangle - \sqrt{2}|001\rangle)/2$, $(|100\rangle + |010\rangle)/\sqrt{2}$, and $(|100\rangle - |010\rangle + \sqrt{2}|001\rangle)/2$. The higher energy states are superpositions of the multiply excited states $|200\rangle$, $|020\rangle$, $|002\rangle$, $|110\rangle$, $|101\rangle$ and $|011\rangle$.}
\label{jjlcjj}
\end{center}
\end{figure}

The lowest three excited states of Eq.~(1) are formed from the subspace spanned by $|100\rangle$, $|010\rangle$ and $|001\rangle$. In our case, $\hbar \omega_2 \approx \hbar \omega_3$, where $\hbar \omega_2$ is the $|0\rangle$ to $|1\rangle$ level spacing for J2. Therefore, for J1 at low bias we expect the first two excited states to be $(|010\rangle \pm |001\rangle)/\sqrt{2}$  with a splitting of $\xi \hbar \omega_3$ (see Fig. 3 for $I_{b1} < 21 \mu$A). We also note the presence of a triple degeneracy point at $I_{b1} \approx 21.15 \mu$A, where the first three excited states make their closest approach. At this bias, the predicted states are $(|100\rangle - |010\rangle - \sqrt{2}|001\rangle)/2$, $(|100\rangle + |010\rangle)/\sqrt{2}$, and $(|100\rangle - |010\rangle + \sqrt{2}|001\rangle)/2$, with corresponding energies of $\hbar \omega_3 (1-\xi/\sqrt{2})$, $\hbar \omega_3$ and $\hbar \omega_3 (1+\xi/\sqrt{2})$. The first and third excited states are entangled states involving the two junctions and the LC resonator, while the second excited state corresponds to an in-phase oscillation of the two junctions that does not couple to the resonator. The higher levels shown in Fig. 3 also agree well with our calculations, and correspond to multiple excitations in all three degrees of freedom.  

The observed avoided crossings at the triple degeneracy point $I_{b1} \approx 21.15 \mu$A exhibit strong coupling with a dimensionless coupling coefficient of $\xi/\sqrt{2} = \sqrt{C_3/(2C + 4C_3)} \approx 0.18$. However, if the LC frequency $\omega_3$ is much greater than either junction frequency, the LC mode can be set to its ground state.  Analysis of this regime using Eq.~(1) shows that the LC mediated interaction arises as a second order perturbation, and can be modeled by a frequency dependent capacitive coupling $\zeta(\omega) = \xi^2/(1-\xi^2 - \omega^2/\omega_3^2)$ when both junctions are tuned to the same frequency $\omega$. This agrees with our previous measurements \cite{Berkley2003s}, which with $\omega/2\pi \approx 5$ GHz and $\omega_3/2\pi \approx 7$ GHz found $\zeta \approx 0.13$, very close to the expected value $\zeta(\omega) = 0.14$. The measurements reported here show that the effective coupling increases from $\xi^2$ to $\xi$ when the junctions are in resonance with the LC mode. Thus if $\xi^2$ were 0.01, then $\xi$ would be 0.1 thereby boosting the coupling strength on resonance by one order of magnitude. Furthermore the off-resonance coupling is proportional to $\xi^4$ when the junction frequencies are detuned from each other and $\omega_3$ is much greater than either junction frequency, hence allowing the dynamic decoupling of each degree of freedom. 

The junctions are separated by almost 1 mm, yet a strong coupling strength between the two can be achieved by tuning them into resonance with a resonator. Based on this resonant coupling method, logic gates can be constructed, similar to those designed \cite{Strauch2003} for capacitive coupling, but with a larger ratio of coupling to decoupling. While the spectroscopic coherence time \cite{Berkley2003B} here is too short ($\sim 2$ ns) for logic gates, it should be possible to increase it using improved qubit isolation, such as an inductive broadband impedance transforming scheme \cite{Martinis2002}.
 
Overall, our results imply that a coupled macroscopic superconducting three-body system comprised of a resonator and two junctions is governed by quantum mechanics. The measurements agree remarkably well with theory, and are spectroscopic evidence for entanglement between all three macroscopic degrees of freedom. The observed strong coupling between two junction qubits separated by almost 1 mm suggests the possibility of achieving controllable coupling between distant qubits.  Finally we note that the fundamental physics of this system was revealed through a spectroscopic technique that can be extended to systems with a large number of qubits.


\begin{acknowledgments}
We thank R. A. Webb for stimulating discussions and helpful advice. We also thank M. A. Gubrud and W. T. Parsons for technical help. We especially thank T. P. Orlando and W. D. Phillips for their insightful comments.  This work was supported by the NSA, the NSF through the QUBIC program under grant number EIA0323261, and the State of Maryland through the Center for Superconductivity Research.

\end{acknowledgments}

\bibliography{scqc2}

\end{document}